\documentclass[aps,pra,reprint,superscriptaddress,nofootinbib]{revtex4-2}

\usepackage{bm}
\usepackage[retainorgcmds]{IEEEtrantools}
\usepackage{graphicx}
\usepackage{mathrsfs}
\usepackage{amsmath}
\usepackage{amsfonts}
\usepackage{amssymb}
\usepackage{color}
\usepackage{times,txfonts}
\usepackage{nicefrac}
\usepackage{ragged2e}
\usepackage{tikz}
\usepackage{enumerate}
\usepackage{mathtools}
\usepackage[normalem]{ulem}
\usepackage[colorlinks=true,linkcolor=blue,urlcolor=blue,citecolor=blue,pdfusetitle]{hyperref}
\usepackage{blindtext}
\usepackage{float} 
\usepackage{scalerel}
\usepackage[normalem]{ulem}
\usepackage[english]{babel}
\usepackage{graphicx}
\usepackage{dcolumn}
\usepackage{bm}
\usepackage{blindtext}
\usepackage{verbatim}
\usepackage{relsize}
\usepackage{mathrsfs}
\usepackage{musicography}
\usepackage{amsmath}
\usepackage{blindtext}
\usepackage{cancel}
\usepackage{physics}
\usepackage{epstopdf}
\usepackage{mathtools}
\usepackage{blindtext}
\usepackage{tensor}
\usepackage{color}
\usepackage[usenames,dvipsnames]{pstricks}
\usepackage{epsfig}
\usepackage{pst-grad} 
\usepackage{pst-plot} 
\usepackage{hyperref}
\usepackage{verbatim}
\usepackage{slashed}
\usepackage{dsfont}
\usepackage{upgreek}
\usepackage{graphicx}

\definecolor{cadmiumgreen}{HTML}{097969}

\newcommand{\victor}[1]{\textcolor{black}{{#1}}}

\begin{document}

\title{Vacuum fluctuations and the renormalized stress–energy tensor on a cone with arbitrary boundary conditions}
\date{\today}

\author{Jo\~ao Paulo  M.  Pitelli}
\email[]{pitelli@unicamp.br}
\affiliation{Departamento de Matem\'atica Aplicada, Universidade Estadual de Campinas,
13083-859 Campinas, S\~ao Paulo, Brazil}

\author{Ricardo A. Mosna}
\email{mosna@unicamp.br}
\affiliation{Instituto de F\'isica Gleb Wataghin, Universidade Estadual de Campinas, 13083-859 Campinas, S\~ao Paulo, Brazil}

\author{Victor Hugo M. Ramos}
\email{vhmarques@usp.br}
\affiliation{Instituto de F\'isica da Universidade de S\~ao Paulo,  05314-970 S\~ao Paulo, Brazil.}

\author{João C. A. Barata}
\email{jbarata@if.usp.br}
\affiliation{Instituto de F\'isica da Universidade de S\~ao Paulo,  05314-970 S\~ao Paulo, Brazil.}

\begin{abstract}

 We analyze the vacuum fluctuations and the stress-energy tensor of a scalar field  of mass $M$ in a conical spacetime, where the topological singularity at the apex requires boundary conditions for the field equation. The necessity of boundary conditions was established by Kay and Studer in the early 1990s, while for $M=0$ stability is achieved only under Dirichlet boundary conditions, and for $M>q$ the field is stable and a localized mode emerges. This mode admits a natural interpretation as a covariant model of an extended particle detector, which allows us to investigate whahow such detectors modify the local vacuum structure. In this framework, the renormalized stress-energy tensor offers a natural way to quantify the influence of the detector on the surrounding spacetime.

\end{abstract}

\maketitle{}

\section{Introduction}

In \(1+2\) dimensions, solutions of Einstein's equations in vacuum \((T_{\mu\nu}=0)\) are locally flat. A central example is the conical spacetime $\mathcal{M}$, which shows how nontrivial topology can give rise to gravitational effects even when local curvature is absent. From a physical perspective, this geometry corresponds to the gravitational 
field generated by a massive point source~\cite{Staruszkiewicz}, and it is 
described by the metric (in units where $c=1$)
\begin{equation}  
ds^2 = -dt^2 + dr^2 + \alpha^2 r^2\, d\theta^2,  
\label{conical metric}
\end{equation}  
with $0 < \alpha < 1$. Here $r \in (0, \infty)$ denotes the radial distance from the source, and the circumference of a circle of radius $r$ is $L = 2\pi\alpha r$. The fact that $L$ is smaller than in $1+2$ Minkowski space reveals the presence 
of an angular deficit
\begin{equation}
\delta = 2\pi(1 - \alpha).
\end{equation}
This angular deficit has direct physical implications: geodesics that are 
initially parallel on opposite sides of the source eventually intersect after 
a finite proper time~\cite{Konkowski1}. Such behavior is particularly relevant 
in the context of gravitational lensing and has been extensively investigated 
in the $(1+3)$-dimensional analog of the cone, the cosmic string spacetime
~\cite{Vilenkin, Sazhin}.

This setup has also been studied in the context of wave scattering. In particular, Deser and Jackiw obtained the scattering amplitude for an incident wave interacting with the conical singularity in Ref.~\cite{Deser}. They took into account the fact that the spacetime is not asymptotically flat by absorbing some undesirable delta functions into the incident-wave definition. 

The same topological structure responsible for such intriguing classical effects also leads to nontrivial phenomena in quantum field theory. In particular, the nontrivial global geometry affects the vacuum structure of quantum fields, modifying correlation functions and altering the behavior of renormalized quantities. These effects arise purely from topology, even in the absence of local curvature~\cite{Konkowski2,Graham,Saharian}.

However, this is not the end of the story. The same nontrivial topology also gives rise to subtle analytical effects. In fact, the corresponding field theory is, in principle, not well posed due to the interaction between the field and the classical singularity at the cone’s apex. This interaction can be effectively modeled by imposing an appropriate boundary condition at $r=0$. In Ref.~\cite{Kay}, Kay and Studer identified the class of boundary conditions that yield a well-defined dynamics for the field. The notion of a “well-defined” or “sensible” dynamics was later formalized by Ishibashi and Wald in Ref.~\cite{WaldII}, and amounts to imposing a small set of physically reasonable requirements: causality, time-translation invariance, and conservation of energy. It was also shown that these sensible dynamics are in  one-to-one correspondence with  positive self-adjoint extensions of the spatial part $A$ of the wave operator
\begin{equation}
\frac{\partial^2\Phi}{\partial t^2}=- A\Phi.
\end{equation}

 Kay and Studer showed that, for a massless scalar field in a conical spacetime, the only positive self-adjoint extension of $A$ is the so-called Friedrichs extension~\cite{Reed}, sometimes also referred to as the generalized Dirichlet boundary condition (since the problem at hand is a singular Sturm-Liouville problem). This restriction stems from the fact that, otherwise, an unstable mode $\sim e^{\pm q t}$ would be present (here $q$ parametrizes the boundary condition, with $q=0$ representing the Friedrichs extension). Although several works discuss this issue (see, e.g., Refs.~\cite{Pitelli1,Lima1,Lima2,Lima3,Lima4}), it remains unclear how to consistently extract two-point functions in the presence of such unstable modes.  Here we show that, once a mass is included and $M>q$, a nontrivial positive extension exists and a normalizable localized mode $\sim \psi(r)e^{-i\sqrt{M^{2}-q^{2}}\,t}$ emerges. \victor{As the mass $M$ becomes negligibly small, our analysis remains valid, but the parameter space for the boundary condition $q$ becomes increasingly restricted. Since the stability of the model requires $q < M$ (to avoid unstable modes), taking the limit $M \to 0$ effectively forces $q \to 0$ (which corresponds to the Dirichlet boundary condition).}

 In Ref.~\cite{Barroso}, one of the present authors considered the  alternative boundary conditions proposed in~\cite{Kay} and showed that their effect on the scattering amplitude is purely additive, in the sense that one recovers the result of Deser and Jackiw~\cite{Deser} plus an extra term arising from the boundary condition. The same idea will be applied here in the construction of the two-point function below. However, unlike in scattering problems, here the localized mode plays a crucial role. \victor{Also in Ref.~\cite{kay2}, Allen {\it et al.} model long range effects around a cosmic string with a finite core through an appropriate boundary condition.}

In Ref.~\cite{Tales1}, a relativistically consistent particle detector was presented. It was obtained from a Lagrangian
\begin{equation}
\mathcal{L} = -\frac{1}{2}\partial_\mu\phi_{\mathrm{D}}\partial^\mu\phi_{\mathrm{D}}
-\frac{1}{2}m_{\mathrm{D}}^2\phi_{\mathrm{D}}^2
+ \bar{\mathcal{L}},
\end{equation}
where $\bar{\mathcal{L}}$ includes the interaction of $\phi_{\mathrm{D}}$ with an additional classical field $\phi_{\mathrm{C}}$, as well as the interaction of $\phi_{\mathrm{C}}$ with a perfect fluid. In this setup, the field $\phi_{\mathrm{C}}$ is necessary to localize (through an interaction potential) one mode of the detector field $\phi_{\mathrm{D}}$, so that it is given by
\begin{equation}
\begin{aligned}
\hat{\phi}_{\mathrm{D}}(x) &= 
 e^{-i\omega_{\mathrm{D}}t} \,\Phi(\bm{x}) \,\hat{a} 
 + e^{i\omega_{\mathrm{D}} t} \,\Phi^{*}(\bm{x}) \,\hat{a}^{\dagger} \\
&\quad + \int \mathrm{d}^{3}\bm{k} \left[ e^{-i\omega_{\bm{k}} t} \,v_{\bm{k}}(\bm{x}) \,\hat{b}_{\bm{k}} 
 + e^{i\omega_{\bm{k}} t} \,v_{\bm{k}}^{*}(\bm{x}) \,\hat{b}_{\bm{k}}^{\dagger} \right],
\end{aligned}
\label{eqdetector}
\end{equation}
where
\begin{equation}
\omega_{\mathrm{D}} = \sqrt{m_{\mathrm{D}}^{2} + \mu}, 
\quad 
\omega_{\bm{k}} = \sqrt{m_{\mathrm{D}}^{2} + \nu_{\bm{k}}};
\label{eq:omegas}
\end{equation}
$\mu > -m_{\mathrm{D}}^2$ is the eigenvalue of the localized mode 
$e^{-i\omega t}\Phi(\bm{x})$, and $\nu_{\bm{k}} > -m_{\mathrm{D}}^2$ represents the continuous spectrum of the generalized eigenfunctions 
$e^{-i\omega_{\bm{k}} t}\varv_{\bm{k}}(\bm{x})$. The perfect fluid is then used to localize $\phi_{\mathrm{C}}$. Within this more elaborate setup, we obtain a stable model for an extended Unruh-DeWitt detector, with $\hat{a}$ and $\hat{a}^\dagger$ in Eq.~(\ref{eqdetector}) playing the role of the ladder operators. \victor{If one is interested in computable transition amplitudes for the detector}, \victor{it is sufficient to} restrict \victor{the} attention to only two states of the detector, namely $|0\rangle$ and its first excited state $\hat{a}^\dagger |0\rangle$, for example. \victor{For many applications, this reduction can be performed consistently \cite{tjoa2020}.}

However, as stated above there is  a simpler way to obtain a normalized mode, without the need for any additional potential or interaction. 
To obtain a localized mode, it suffices to impose at the apex a non-Dirichlet boundary condition (self-adjoint extension parameter $q\ne0$) and work in the massive regime $M>q$.
This choice yields a much simpler setup for modeling an extended Unruh-DeWitt detector. This is our main motivation for calculating the renormalized stress-energy tensor for massive fields on the cone.  This calculation can be used to quantify how this relativistically consistent detector affects the surrounding spacetime. \victor{Therefore, we focus solely on the study of the detector model itself, without considering its interaction to a target field.}

\section{Scalar Field Dynamics}\label{sec:field_dynamics}

In this section, we follow the analysis of Ref.~\cite{Kay} and consider a minimally coupled massive real scalar field $\Psi$ satisfying (in units where $c=\hbar=1$)
\begin{equation}\label{klein-gordon}
(\Box_g - M^2)\Psi = 0,
\end{equation}
where $\Box_g$ is the d'Alembert operator for the conical spacetime metric~(\ref{conical metric}). In view of the cylindrical symmetry of the metric, we consider the ansatz
\begin{equation}\label{ansatz}
\Psi(t,r,\theta) = e^{-i\omega t} e^{i n \theta} R_{\omega n}(r),
\end{equation}
where the functions $R_{\omega n}$ satisfy the singular Sturm-Liouville problem
\begin{equation}\label{sturm}
R''_{\omega n}(r) + \frac{1}{r} R'_{\omega n}(r)
+ \left[(\omega^2 - M^2) - \frac{n^2}{\alpha^2 r^2}\right] R_{\omega n}(r) = 0.
\end{equation}
The general solution of Eq.~(\ref{sturm}) is
\begin{equation}\label{solution-sturm}
R_{\omega n}(r) =
\begin{cases}
N_n J_{\frac{|n|}{\alpha}}(\lambda r), & n \neq 0, \\[4pt]
N_0 \left[J_0(\lambda r) + \beta(\lambda) Y_0(\lambda r)\right], & n = 0,
\end{cases}
\end{equation}
where $J_\nu(x)$ and $Y_\nu(x)$ are the Bessel functions of the first and second kinds, respectively, and $\lambda = \sqrt{\omega^2 - M^2}$ is the spectral parameter. Furthermore, the coefficients $N_n$ and $\beta(\lambda)$ will be later defined by the imposition of a proper normalization and boundary conditions at the apex, respectively.

Regarding the profile of the solutions, we note that when $n \neq 0$, the strong repulsive potential $-n^2/\alpha^2 r^2$ effectively decouples the field from the singularity. Mathematically, this means that $Y_{\frac{|n|}{\alpha}}(\lambda r)$ is not square-integrable near $r = 0$ for $n \neq 0$, hence the equation is in the limit-point case. On the other hand, for $n = 0$ the equation is on the limit-circle case, i.e. $Y_0(x)$ is also square-integrable near $x = 0$ and cannot, in principle, be excluded.

The factor $\beta(\lambda)$ is given by
\begin{equation}
\beta(\lambda) = \frac{\pi}{2\log\left(\frac{q}{\lambda}\right)}
\end{equation}
and is introduced to ensure that $R_{\omega 0}$ satisfies the boundary condition corresponding to the self-adjoint extensions of the operator in Eq.~(\ref{sturm}) for $n = 0$ (see Ref.~\cite{Kay} for details):
\begin{equation}\label{bc}
\begin{aligned}
&\lim_{r \to 0} \left\{ \left[ \ln\left(\frac{r}{Q}\right) r\frac{d}{dr} - 1 \right] R_{\omega 0}(r) \right\}, \quad Q \in (0,\infty),\\
&\lim_{r \to 0} r\frac{d}{dr} R_{\omega 0}(\lambda r) = 0, \quad Q = 0.
\end{aligned}
\end{equation}
The choice of $Q$ parametrizes the self-adjoint extension of the corresponding radial Sturm-Liouville operator, and it is convenient to recast it as
\begin{equation}
q = 2 e^{-\gamma} Q^{-1},
\end{equation}
with $ q = 0 $ and $q = \infty$ identified. \victor{Indeed, this identification is justified because the spatial part of the wave operator, $A^{(q)}$, converges to the same operator $A^{(0)}$ in both limits ($q \to \infty$ and $q \to 0$). Consequently, the parameter space effectively ``loops'' back on itself, compactifying the real line into a circle rather than remaining an open interval \cite{Kay}.} Under these considerations we arrive at Eq.~(\ref{solution-sturm}), and characterize the most general solutions satisfying the boundary conditions above. By inserting the radial solutions into Eq.~(\ref{ansatz}), the continuous modes take the form
\begin{equation}\label{eq:continum-modes}
u_{\omega n}(t,r,\theta) = e^{-i \omega t} e^{i n \theta}
\begin{cases}
N_n J_{\frac{|n|}{\alpha}}(\lambda r), & n \neq 0, \\[4pt]
N_0 \left[J_0(\lambda r) + \beta(\lambda) Y_0(\lambda r)\right], & n = 0,
\end{cases}
\end{equation}
where the normalization factors $N_n$, for $n = 0,1,2,\dots$, are fixed through the Klein-Gordon inner product,
\begin{equation}
\langle u_{\omega' n'} , u_{\omega n} \rangle = \delta(\omega - \omega') \delta_{n n'},
\end{equation}
A straightforward calculation yields
\begin{equation}
N_n =
\begin{cases}
\dfrac{1}{\sqrt{4\pi\alpha}}, & n \neq 0, \\[6pt]
\dfrac{1}{\sqrt{4\pi\alpha}} \dfrac{1}{\sqrt{1 + \beta^2(\lambda)}}, & n = 0.
\end{cases}
\end{equation}

The continuous modes are not the whole story. There exists one discrete bound mode with positive frequency satisfying Eq.~(\ref{sturm}) and the boundary condition~(\ref{bc}), given in normalized form by
\begin{equation}\label{eq:bound-mode}
\Psi_{\mathrm{bound}}(t,r,\theta) =
\frac{q}{\sqrt{2\pi\alpha}} \frac{K_0(q r)}{(M^2 - q^2)^{1/4}}
e^{-i \sqrt{M^2 - q^2}\, t},
\end{equation}
where $K_0$ is the modified Bessel function of the second kind. We notice that the radial solution $K_0(q r)$ is square-integrable for $0<r<\infty$. As a result, $\Psi_{\mathrm{bound}}$ characterizes a localized mode which is physically sensible and suitable for modeling an extended particle detector. 

The quantum field $\hat{\Psi}(t,r,\theta)$ follows directly from the expansion on the complete set of modes in Eqs.~(\ref{eq:continum-modes}) and (\ref{eq:bound-mode}), and it takes the form
\begin{equation}\begin{aligned}
\hat{\Psi}(t,r,\theta)&=\hat{a}_{\textrm{bound}}\Psi_{\textrm{bound}}+\hat{a}^\dagger_{\textrm{bound}}\Psi^{\ast}_{\textrm{bound}}\\&+\sum_{n=-\infty}^{\infty}\int_M^\infty d\omega{\left(\hat{b}_{\omega n}u_{\omega n}+\hat{b}_{\omega n}^\dagger u^\ast_{\omega n}\right)},
\end{aligned}\end{equation}
where the bound solution translates into the discrete mode. The creation and annihilation operators $\hat{a}_{\textrm{bound}}$, $\hat{a}^\dagger_{\textrm{bound}}$, $\hat{b}_{\omega n}$, $\hat{b}^\dagger_{\omega n}$ satisfy the canonical commutation relations \begin{align}
    \left[ \hat{a}_{\textrm{bound}},\hat{a}_{\textrm{bound}}^\dagger \right] & = 1, \\
    \left[ \hat{b}_{\omega n},\hat{b}_{\omega' n'}^\dagger \right] & = \delta(\omega-\omega') \delta_{n n'}.
\end{align} The annihilation operators define the vacuum state $|0\rangle$ via $\hat{a}_{\textrm{bound}} | 0 \rangle = 0 $ and $\hat{b}_{\omega n}| 0\rangle = 0 $ for all $\omega \in [M,\infty)$ and $n\in\mathbb{Z}$. They also define the vacuum of each sector: the state $|0_{\textrm{bound}}\rangle$, satisfying 
$\hat{a}_{\textrm{bound}} |0_{\textrm{bound}}\rangle = 0$, represents the vacuum of the discrete (bound) mode, while the state $|0_{\textrm{cont}}\rangle$, satisfying 
$\hat{b}_{\omega \ell m} |0_{\textrm{cont}}\rangle = 0$, represents the vacuum of the continuum sector. Consequently, the complete vacuum decomposes into a tensor product $|0\rangle = |0\rangle_{\textrm{bound}} \otimes |0\rangle_{\textrm{cont}}$.

\vspace{5mm}

\section{Two-Point Function}\label{Sec:TPF}

The study of quantum fluctuations and the associated energy-momentum content naturally starts from the explicit form of the two-point correlation function~\footnote{In fact, the symmetric expression 
\[
\frac{1}{2}G^{(1)}(\mathsf{x},\mathsf{x}')
=\frac{1}{2}\langle 0|\,\Psi(\mathsf{x})\Psi(\mathsf{x}')+\Psi(\mathsf{x}')\Psi(\mathsf{x})\,|0\rangle
\]
should in principle be used. This becomes relevant when computing 
$\langle T_{\alpha\beta}\rangle$ with $\alpha\neq\beta$, as using a nonsymmetric 
correlator would generally yield a complex-valued renormalized stress-energy tensor. 
However, by symmetry arguments of the spacetime, we know that the only nonvanishing 
components of $\langle T_{\alpha\beta}\rangle$ are those with $\alpha=\beta$. 
Therefore, using either $G^{+}(\mathsf{x},\mathsf{x}')$ or 
$\frac{1}{2}G^{(1)}(\mathsf{x},\mathsf{x}')$ leads to the same final result.} 
\begin{equation}
G^{+}(\mathsf{x},\mathsf{x}')=\langle 0|\Psi(\mathsf{x})\Psi(\mathsf{x}')|0\rangle
\end{equation} 
associated  with the vacuum state of a massive scalar field in the conical spacetime~(\ref{conical metric}). 
The function is obtained as a mode expansion on both the discrete and the continuum sectors. The former is described by the modes in Eq.~(\ref{eq:bound-mode}), and the latter by the solutions in Eq.~(\ref{eq:continum-modes}). The resulting expansion takes the form
\begin{equation}
G^+(\mathsf{x},\mathsf{x}')=\Psi_\textrm{bound}(x)\Psi_\textrm{bound}^\ast(x')+ \sum_{n\in\mathbb{Z}} \int_{M}^\infty d\omega \, u_{\omega n}(\mathsf{x})\,u^\ast_{\omega n}(\mathsf{x}'). 
\end{equation}

This expression is not the most convenient representation of 
$G^{+}(\mathsf{x},\mathsf{x}')$ when dealing with non-Dirichlet boundary 
conditions. It is often preferable to separate $G^{+}$ into a purely Dirichlet 
contribution and an additional term induced by the nontrivial boundary 
condition. Since the boundary conditions affect exclusively the axisymmetric sector ($n=0$), Green's function can be naturally decomposed into three distinct contributions: (i) the discrete bound state modes arising from the self-adjoint extension of the radial operator [$G_{\textrm{bound}}(x,x')$], (ii) the continuous  mode in the $\ell=0$ sector modified by the non-Dirichlet boundary conditions [$G_{\textrm{bc}}(x,x')$], and (iii) the modes corresponding to the Dirichlet boundary condition which remain unaffected by the chosen extension [$G_{\textrm{Dirichlet}}(x,x')$].  This separation can be implemented by adding and subtracting the azimuthal $n=0$ mode, assigning the added term to $G_{\textrm{Dirichlet}}(x,x')$, while the remaining piece is absorbed into $G_{\textrm{bc}}(x,x')$, ensuring that the latter vanishes smoothly in the Dirichlet limit. Hence, the resulting two-point function is \begin{equation}
\begin{aligned}
G^{+}(\mathsf{x},\mathsf{x}')\equiv G^{+}_{\textrm{Dirichlet}}(\mathsf{x},\mathsf{x}')
+ G^{+}_{\textrm{bc}}(\mathsf{x},\mathsf{x}')
+ G^{+}_{\textrm{bound}}(\mathsf{x},\mathsf{x}'),
\end{aligned}
\label{decomposition}
\end{equation} where each contribution is explicitly given by 
\begin{widetext}
\begin{equation}
\begin{aligned}
&G^{+}_{\textrm{Dirichlet}}(\mathsf{x},\mathsf{x}')=\frac{1}{4\pi\alpha}\sum_{n=-\infty}^{\infty}\int_0^\infty \textrm{d}\lambda\frac{\lambda}{\omega} J_{\frac{|n|}{\alpha}}(\lambda r)J_{\frac{|n|}{\alpha}}(\lambda r')e^{in \Delta\theta}e^{-i\omega \Delta t}, \\
&G^+_{\textrm{bc}}(\mathsf{x},\mathsf{x}')=\frac{1}{4\pi\alpha}\int_0^\infty \textrm{d}\lambda\frac{\lambda}{\omega}\frac{\beta(\lambda)}{1+\beta^2(\lambda)}\Bigg\{\beta(\lambda)\Big[-J_0(\lambda r)J_0(\lambda r')+Y_0(\lambda r)Y_0(\lambda r')\Big]+\Big[J_0(\lambda r)J_0(\lambda r')+Y_0(\lambda r)Y_0(\lambda r')\Big]\Bigg\}e^{-i\omega\Delta t},\\
&G^+_{\textrm{bound}}(\mathsf{x},\mathsf{x}')= \frac{q^2}{2\pi \alpha}\frac{K_0(q r)K_0(q r')}{(M^2-q^2)^{1/2}}e^{-i\sqrt{M^2-q^2}\Delta t}.
\end{aligned}
\end{equation}
\end{widetext}

 We follow the analysis by considering each contribution separately, beginning with the Dirichlet sector. To evaluate $G^{+}_{\textrm{Dirichlet}}$, we perform a Wick rotation $\Delta \tau = i \Delta t$ and use the well-known identity \begin{equation}\label{Wick}
\frac{e^{\omega \Delta\tau}}{\omega}
= \frac{2}{\sqrt{\pi}}\int_0^\infty \mathrm{d}s \,
e^{-\lambda^2 s^2}\, e^{-\tfrac{\Delta \tau^2}{4s^2}},
\end{equation}
together with the integral identity from ~\cite{Gradshteyn},
\begin{equation}\label{preidentity}
\int_0^{\infty}\!\mathrm{d}\lambda\, \lambda \,
J_{\tfrac{|n|}{\alpha}}(\lambda r)\,
J_{\tfrac{|n|}{\alpha}}(\lambda r')\, e^{-\omega^2 s^2}
= \frac{e^{-\tfrac{(r^2+r'^2)}{4s^2}}}{2s^2}\,
I_{\tfrac{|n|}{\alpha}}\!\left(\frac{r r'}{2s^2}\right).
\end{equation}
Combining these relations, we obtain 
\begin{multline}\label{G com I}
G^{+}_{\textrm{Dirichlet}} (\mathsf{x},\mathsf{x}')
= \frac{1}{4\alpha \pi^{3/2}\,}
\int_0^{\infty}\!\mathrm{d}s \left[ \dfrac{e^{-\tfrac{(r^2+r'^2)}{4s^2}}}{s^2} e^{-M^2 s^2 - \tfrac{\Delta\tau^2}{4s^2}} \right.
\\ \left. \times \sum_{n=0}^{\infty} 
I_{\tfrac{|n|}{\alpha}}\!\left(\frac{r r'}{2s^2}\right) e^{i n \Delta\theta}\right].
\end{multline} The remaining summation over $n$ can be further simplified by employing the identity from Ref.~\cite{Mota}:
\begin{equation}\label{identity}
\begin{aligned}
\sum_{n=0}^{\infty} I_{\tfrac{|n|}{\alpha}}(y) e^{i n \Delta\theta}
&= \sum_{n=-\left[\tfrac{1}{2\alpha}\right]}^{\left[\tfrac{1}{2\alpha}\right]}
\alpha \, e^{\,y \cos(2\pi \alpha n - \alpha \Delta \theta)} \\
&\quad - \frac{1}{2\pi}\sum_{j=\pm}
\int_0^\infty\!\mathrm{d}\xi \,
\frac{\sin\!\left[\tfrac{j\alpha \Delta\theta + \pi}{\alpha}\right]
\, e^{-y \cosh \xi}}
{\cosh\!\left(\tfrac{\xi}{\alpha}\right)
- \cos\!\left(\tfrac{j\alpha \Delta\theta + \pi}{\alpha}\right)}.
\end{aligned}
\end{equation}
Here, $[1/(2\alpha)]$ denotes the integer part of $1/(2\alpha)$, and in the special case where $1/(2\alpha)$ is a positive integer, the summation over $n$ in Eq.~(\ref{identity}) must be taken with an overall factor of $1/2$.
Substituting Eq.~(\ref{identity}) into Eq.~(\ref{G com I}), we finally obtain
\begin{equation}\label{eq:GreensDirichlet}
\begin{aligned}
G^+_{\textrm{Dirichlet}} (\mathsf{x},\mathsf{x}')
&= \sum_{n=-\left[\tfrac{1}{2\alpha}\right]}^{\left[\tfrac{1}{2\alpha}\right]}
\frac{1}{4\pi}\,
\frac{e^{-M \sqrt{2\sigma_n}}}{\sqrt{2\sigma_n}} \\[6pt]
&\quad - \frac{1}{8\pi^2 \alpha}
\sum_{j=\pm} \int_0^\infty\!\mathrm{d}\xi \,
\frac{\sin\!\left[\tfrac{j\alpha \Delta\theta + \pi}{\alpha}\right] \,
\tfrac{e^{-M\sqrt{2\sigma_\xi}}}{\sqrt{2\sigma_\xi}}}
{\cosh\!\left(\tfrac{\xi}{\alpha}\right)
- \cos\!\left(\tfrac{j\alpha \Delta\theta + \pi}{\alpha}\right)},
\end{aligned}
\end{equation}
with
\begin{equation}
\begin{aligned}
\sigma_n &= \tfrac{1}{2}\Big[r^2 + r'^2 - \Delta t^2 
- 2 r r' \cos(\alpha \Delta\theta - 2\pi n \alpha)\Big], \\[6pt]
\sigma_\xi &= \tfrac{1}{2}\Big[r^2 + r'^2 - \Delta t^2 
+ 2 r r' \cosh \xi\Big].
\end{aligned}
\end{equation} It follows from this representation that $G^+_{\textrm{Dirichlet}}$ exhibits the characteristic short-distance singularities of the theory, which must be consistently regularized by subtraction. In particular, the relevant Hadamard parametrix~\cite{Decanini} in $1+2$ dimensions takes the form
\begin{equation}\label{eq:parametrix}
G_H(\mathsf{x},\mathsf{x}')=\frac{1}{4\pi \sqrt{2\sigma_0}}\,U(\mathsf{x},\mathsf{x}')+W(\mathsf{x},\mathsf{x}'),
\end{equation}
where $W(\mathsf{x},\mathsf{x}')$ denotes the smooth state-dependent contribution and the function $U(\mathsf{x},\mathsf{x})$ is entirely determined by the local geometry and the field dynamics. Specifically, the latter admits an expansion in powers of the squared geodesic interval $\sigma_0 (\mathsf{x},\mathsf{x}')$,
\begin{equation}
U(\mathsf{x},\mathsf{x}')
=\sum_{n=0}^{\infty} U_n(\mathsf{x},\mathsf{x}') \, \sigma_0^n(\mathsf{x},\mathsf{x}'),
\end{equation} where the Hadamard coefficients $U_{n}(\mathsf{x},\mathsf{x}')$ satisfy the recurrence relation (see Ref.~\cite{Decanini} for details) \begin{equation}\label{eq:recurrence}
\begin{aligned}
&(n+1)(2n+1)U_{n+1}
+(2n+1)U_{n+1;\mu}\,\sigma^{;\mu} \\
&\quad -(2n+1)U_{n+1;\mu}\,\Delta^{-1/2}\Delta^{1/2}_{;\mu}\,\sigma^{;\mu} \\
&\quad +(\square_x-M^2)U_{n}=0,
\qquad n=1,2,\dots \, .
\end{aligned}
\end{equation} with the boundary condition $U_0=\Delta^{1/2}$. Furthermore, the biscalar function $\Delta(\mathsf{x},\mathsf{x}')$ denotes the Van Vleck-Morette determinant, 
which is defined by
\begin{equation}
\Delta(\mathsf{x},\mathsf{x}')=-[-g(\mathsf{x})]^{-1/2}
\det[-\sigma_{0;\mu\nu'}(\mathsf{x},\mathsf{x}')]
[-g(\mathsf{x}')]^{-1/2}
\end{equation} and satisfies the boundary condition \begin{align}
    \lim_{\mathsf{x}'\to\mathsf{x}} \Delta(\mathsf{x},\mathsf{x}')=1.
\end{align} In fact, a direct computation shows that in the present case $\Delta(\mathsf{x},\mathsf{x}')=1$ for every $\mathsf{x},\mathsf{x}'\in\mathcal{M}$.

We now determine the Hadamard coefficients $U_n (\mathsf{x},\mathsf{x}')$ by isolating the short-distance singular structure of $G^+_{\textrm{Dirichlet}}$. This can be achieved by expanding this contribution at the singularity and then comparing the result with the Hadamard parametrix in Eq.~(\ref{eq:parametrix}). Since the expansion takes the form

\begin{multline}
    \frac{e^{-M\sqrt{2\sigma_0}}}{4\pi\sqrt{2\sigma_0}}
= \frac{1}{4\pi\sqrt{2\sigma_0}}
\Bigg(1+\frac{M^2}{2!}(2\sigma_0)+\frac{M^4}{4!}(2\sigma_0)^2+\cdots\Bigg) \\
\quad - \frac{1}{4\pi\sqrt{2\sigma_0}}
\Bigg(M\sqrt{2\sigma_0}+\frac{M^3}{3!}(2\sigma_0)^{3/2}
+\frac{M^5}{5!}(2\sigma_0)^{5/2}+\cdots\Bigg),
\end{multline} we renormalize the two-point function $G^{+}_{\textrm{Dirichlet}}$ by subtracting the singular part identified as
\begin{equation}\label{eq:GreenSingular}
\begin{aligned}
G_\textrm{sing}(\mathsf{x},\mathsf{x}')
&=\frac{1}{4\pi \sqrt{2\sigma_0}}U(\mathsf{x},\mathsf{x}') \\
&=\frac{1}{4\pi\sqrt{2\sigma_0}}
\Bigg(1+\frac{M^2}{2!}(2\sigma_0)+\frac{M^4}{4!}(2\sigma_0)^2+\cdots\Bigg).
\end{aligned}
\end{equation} where the geometric coefficients are characterized by this expansion as
$U_0(\mathsf{x},\mathsf{x}')=1$, 
$U_1(\mathsf{x},\mathsf{x}')=M^2$, 
$U_2(\mathsf{x},\mathsf{x}')=M^4/3$, and the subsequent coefficients determined accordingly. Moreover, these Hadamard coefficients satisfy the expected recursive relation in Eq.~(\ref{eq:recurrence}).

 After subtracting the singular contribution in Eq.~(\ref{eq:GreenSingular}) from the original Green's function in Eq.~(\ref{eq:GreensDirichlet}), we obtain the renormalized two-point function
\begin{equation}
\begin{aligned}
G^{+(\textrm{ren})}_{\textrm{Dirichlet}} (\mathsf{x},\mathsf{x}') &=-\frac{1}{4\pi\sqrt{2\sigma_0}}\left(M\sqrt{2\sigma_0}+\frac{M^3}{3!}(2\sigma_0)^{3/2}+\cdots\right)\\
&+\sum_{n=-\left[\tfrac{1}{2\alpha}\right]}^{-1}\frac{1}{4\pi}\frac{e^{-M \sqrt{2\sigma_n}}}{\sqrt{2\sigma_n}}+\sum_{n=1}^{\left[\tfrac{1}{2\alpha}\right]}\frac{1}{4\pi}\frac{e^{-M \sqrt{2\sigma_n}}}{\sqrt{2\sigma_n}}\\
&-\frac{1}{8\pi^2\alpha}\sum_{j=\pm}\int_0^\infty\textrm{d}\xi\,\frac{\sin\!\left[\tfrac{j\alpha \Delta\theta+\pi}{\alpha}\right]\tfrac{e^{-M\sqrt{2\sigma_\xi}}}{\sqrt{2\sigma_\xi}}}{\cosh\!\left(\tfrac{\xi}{\alpha}\right)-\cos\!\left(\tfrac{j\alpha\Delta\theta+\pi}{\alpha}\right)}
\\&=-\frac{M}{4\pi}+G_\textrm{extra}(\mathsf{x},\mathsf{x}'),
\end{aligned}
\label{Grenorm}
\end{equation}
with 
\begin{equation}\begin{aligned}
G_{\textrm{extra}}(\mathsf{x},\mathsf{x}')&=-\frac{M^3}{12\pi}\sigma_0-\frac{M^5}{120\pi}\sigma_0^2+\cdots    \\&+\sum_{n=-\left[\tfrac{1}{2\alpha}\right]}^{-1}\frac{1}{4\pi}\frac{e^{-M \sqrt{2\sigma_n}}}{\sqrt{2\sigma_n}}+\sum_{n=1}^{\left[\tfrac{1}{2\alpha}\right]}\frac{1}{4\pi}\frac{e^{-M \sqrt{2\sigma_n}}}{\sqrt{2\sigma_n}}\\
&-\frac{1}{8\pi^2\alpha}\sum_{j=\pm}\int_0^\infty\textrm{d}\xi\,\frac{\sin\!\left[\tfrac{j\alpha \Delta\theta+\pi}{\alpha}\right]\tfrac{e^{-M\sqrt{2\sigma_\xi}}}{\sqrt{2\sigma_\xi}}}{\cosh\!\left(\tfrac{\xi}{\alpha}\right)-\cos\!\left(\tfrac{j\alpha\Delta\theta+\pi}{\alpha}\right)}
\end{aligned}\end{equation}
being convergent in the coincidence limit $\mathsf{x}'\to\mathsf{x}$.

The remaining contributions in Eq.~(\ref{decomposition}) arise from the 
choice of nontrivial boundary conditions. Unlike the Dirichlet part, these 
terms are regular in the coincidence limit $\mathsf{x}\to \mathsf{x}'$, 
and therefore do not require any additional subtraction. 
They encode the specific modifications to the two-point function induced 
by the boundary condition, including both the continuous and bound-state 
sectors. For the contribution $G^{+}_{\textrm{bc}}$, the integral expression cannot be 
simplified into a closed analytic form. Consequently, its evaluation must be 
performed numerically.

\section{Vacuum Fluctuations}

\begin{figure}[h!]
\includegraphics[width=\linewidth]{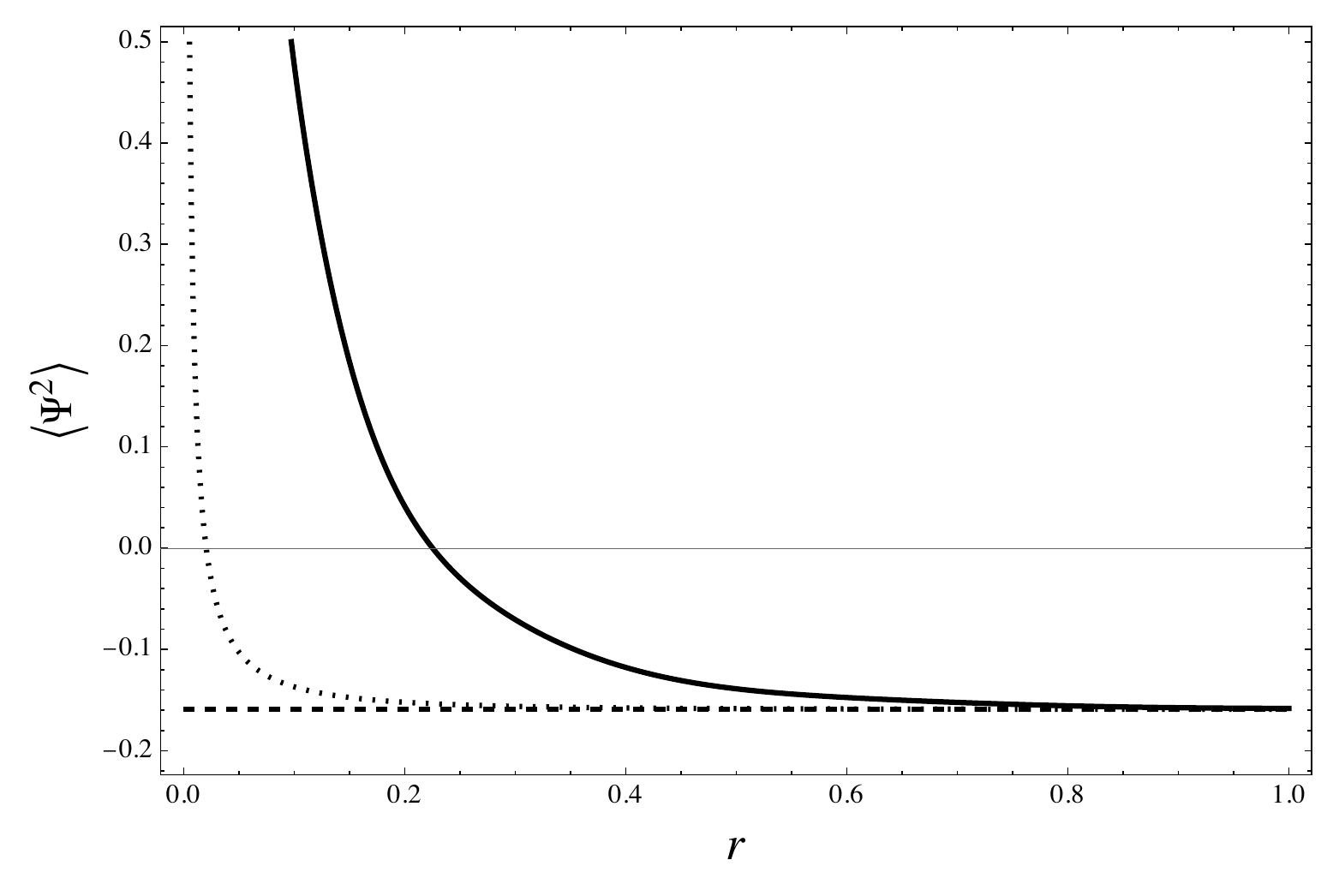}
\caption{$\langle \Psi^2\rangle_{\textrm{Dirichlet}}$ (dotted line) as a function of $r$ for $M=2$ and $\langle \Psi^2\rangle$ (solid line) for $\alpha=0.9$, $q=1$ and $M=2$. As $r\to\infty$, we recover the Minkowski value $-M/(4\pi)$ (dashed line); as $r\to 0$, the boundary-condition contribution dominates the Dirichlet contribution.} 
\label{fig1}
\end{figure}

Having obtained the renormalized two-point function, we now turn to the evaluation of local observables. The first quantity of interest is the vacuum expectation value of the squared field operator, which encodes the fluctuations of the quantum field in the conical background. For the Dirichlet contribution, the renormalized fluctuations are given by
\begin{equation}\label{eq:Psi2Dirichlet}
\langle \Psi^2\rangle_{\textrm{Dirichlet}} =\lim_{\mathsf{x}'\to\mathsf{x}}{G^{+(ren)}_{\textrm{Dirichlet}}(\mathsf{x},\mathsf{x}')}=-\frac{M}{4\pi}+\langle \Psi^2\rangle_\textrm{extra},
\end{equation}
where  the term $-\frac{M}{4\pi}$  agrees with $\langle \Psi^2\rangle$ for a massive scalar field in $(1+2)$-dimensional Minkowski spacetime~\cite{Ferreira}. In this paper, we consider only $\alpha>1/2$. Thus,
\begin{equation}\label{eq:psi2extra}
\langle \Psi^2\rangle_\textrm{extra}=-\frac{1}{4\pi^2\alpha}\int_0^{\infty}\textrm{d}\xi\frac{\sin{\left(\frac{\pi}{\alpha}\right)}\frac{e^{-M r\sqrt{(1+\cosh{\xi})}}}{r\sqrt{(1+\cosh{\xi})}}}{\cosh{\left(\frac{\xi}{\alpha}\right)}-\cos{\left(\frac{\pi}{\alpha}\right)}}
\end{equation}
This function has no short-distance singularities since the divergent part of the two-point function $G^+$ has already been subtracted consistently. Furthermore, the integral in Eq.~(\ref{eq:psi2extra}) is convergent and can be treated numerically. The plot of $\langle \Psi^2\rangle_{\textrm{Dirichlet}}$ as a function of $r$ can be found in Fig.~\ref{fig1}.

There still remain two  regular components of $\langle \Psi \rangle $ originated from Green's function of the discrete mode and the boundary condition, namely $G^+_{\textrm{bound}}$ and $G^+_{\textrm{bc}}$.  Both contributions are obtained by the same coincidence limit procedure in Eq.~(\ref{eq:Psi2Dirichlet}), and they, respectively, take the form  
\begin{equation}
\langle \Psi^2\rangle_{\textrm{bound}}=\frac{q^2}{2\pi\alpha}\frac{K_0(q r)^2}{\sqrt{M^2-q^2}}
\end{equation}
and
\begin{equation}\begin{aligned}
\langle \Psi^2\rangle_{\textrm{bc}}&=\frac{1}{4\pi\alpha}\int_0^\infty\textrm{d}\lambda \frac{\lambda}{\omega}\frac{1}{1+\beta^2(\lambda)}\Big\{\beta^2(\lambda)\left[-J_0^2(\lambda r)+Y_0^2(\lambda r)\right]\\&+\beta(\lambda)\left[J_0^2(\lambda r)+Y_0^2(\lambda r)\right]\Big\}.
\end{aligned}\end{equation}
The full vacuum fluctuation of the field squared is therefore given by\begin{align}
	\langle \Psi^2 \rangle = \langle \Psi^2\rangle_{\textrm{Dirichlet}}+\langle \Psi^2\rangle_{\textrm{bound}} + \langle \Psi^2\rangle_{\textrm{bc}}.
\end{align} The boundary condition contribution $\langle \Psi^2 \rangle_{\textrm{bc}}$ must be calculated numerically  as the integral lacks a closed-form expression. In Fig.~\ref{fig1} we plot
$\langle \Psi^2\rangle$ as a function of $r$ for $M=2$ and $q=1$. We  separate the Dirichlet contribution,  $\langle \Psi^2\rangle_{\textrm{Dirichlet}}$ (dotted line) from the full result, $\langle \Psi^2\rangle$ (solid line). We see that  the contributions coming solely from the nontrivial boundary condition approach zero as $r\to \infty$, when far from the singularity. 

\section{Renormalized stress-energy tensor}
\label{RSET}

 In the present conical background, the renormalized stress-energy tensor captures the local modifications arising from nontrivial boundary conditions at the apex. As in the case of the vacuum fluctuations of $\langle \Psi^2 \rangle$, the full stress-energy tensor can be naturally decomposed into three contributions: the Dirichlet sector $\langle T_{\mu\nu} \rangle_{\textrm{Dirichlet}}$, which reflects the standard cone geometry; the bound-state sector $\langle T_{\mu\nu} \rangle_{\textrm{bound}}$, associated with the localized mode supported for $M > q$; and the boundary-condition sector $\langle T_{\mu\nu} \rangle_{\textrm{bc}}$. This decomposition will allow us to study the effects induced by the boundary condition, and to investigate how the localized mode modifies the local stress distribution.  

Following the same order of the previous sections, we start by considering the contribution due to the Dirichlet boundary condition.  From Ref.~\cite{Decanini}, by defining 
\begin{equation}\begin{aligned}
W_{\textrm{Dirichlet}}(\mathsf{x},\mathsf{x}')&=G^{+(\textrm{ren})}_{\textrm{Dirichlet}}(\mathsf{x},\mathsf{x}'),\\
W_{\textrm{Dirichlet}}&=\lim_{\mathsf{x}\to\mathsf{x}'}W_{\textrm{Dirichlet}}(\mathsf{x},\mathsf{x}'),\\
W_{{\textrm{Dirichlet}}\,\mu\nu}&=\lim_{\mathsf{x}\to\mathsf{x}'}\nabla_\mu\nabla_\nu W_{\textrm{Dirichlet}}(\mathsf{x},\mathsf{x}'),
\end{aligned}\end{equation}
  we obtain
\begin{align}
\langle T_{\mu\nu}\rangle_{\textrm{Dirichlet}}=&-W_{{\textrm{Dirichlet}}\,\mu\nu}+\frac{1}{2}W_{\textrm{Dirichlet};\mu\nu} \nonumber \\ & \qquad -\frac{1}{4}g_{\mu\nu}\square W_{\textrm{Dirichlet}}+\Theta_{\mu\nu}.
\end{align}
The tensor $\Theta^{\mu\nu}$ in the above equation is purely geometric   and  does not affect the conservation of $T_{\mu\nu}$. It represents an ambiguity in the renormalization procedure and satisfies (in odd dimensions)
\begin{equation}
\Theta^{\mu\nu}_{\phantom{\mu\nu};\nu}=0.
\end{equation}
In particular, in $1+2$ dimensions, it is given by~\cite{Decanini}
\begin{equation}
\Theta_{\mu\nu} = A M^{3} g_{\mu\nu} + B M^{2} \left[ R_{\mu\nu} - \tfrac{1}{2} R g_{\mu\nu} \right].
\end{equation}
In our case, since $R_{\mu\nu}=0$, we have simply $\Theta_{\mu\nu}=AM^3g_{\mu\nu}$. In what follows, our choice for the constant $A$ will be such that $\langle T_{\mu\nu}\rangle_{\textrm{Dirichlet}}=0$ for $\alpha=1$ (when we recover Minkowski spacetime). This choice coincides with the usual prescription of normal ordering in flat spacetime.

\begin{figure}[htb!]
\centering
    \includegraphics[width=\linewidth]{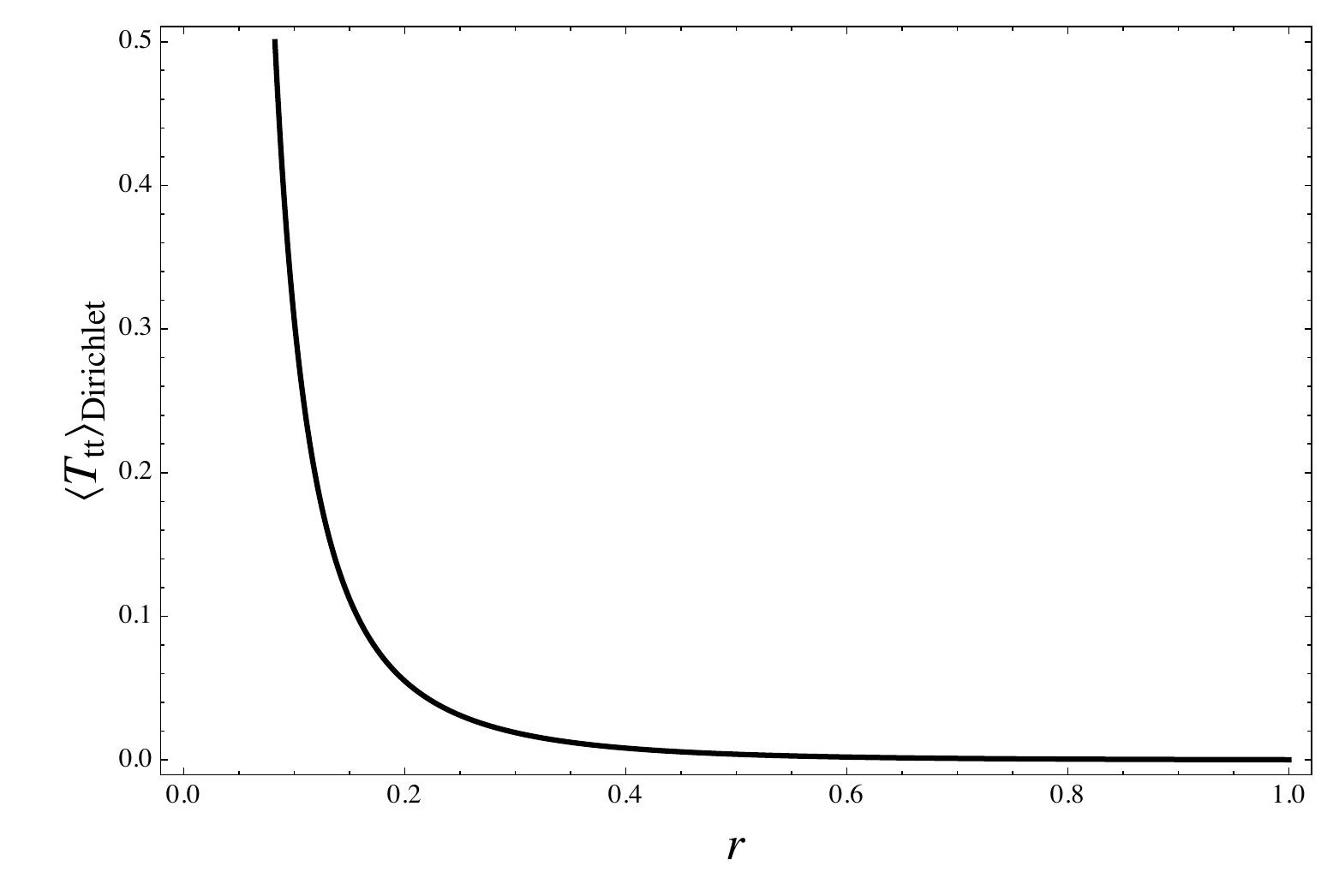}
    \includegraphics[width=\linewidth]{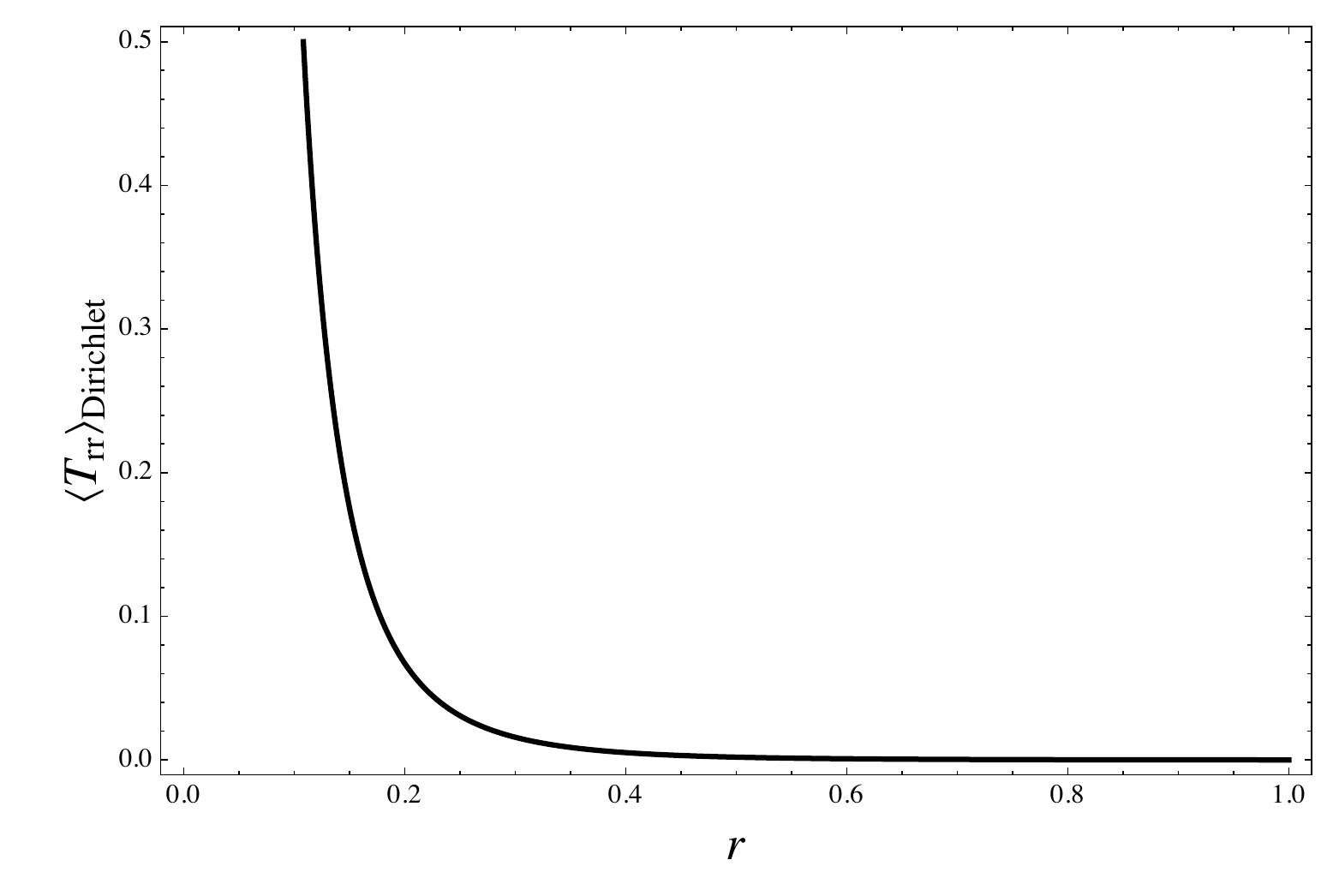}
    \includegraphics[width=\linewidth]{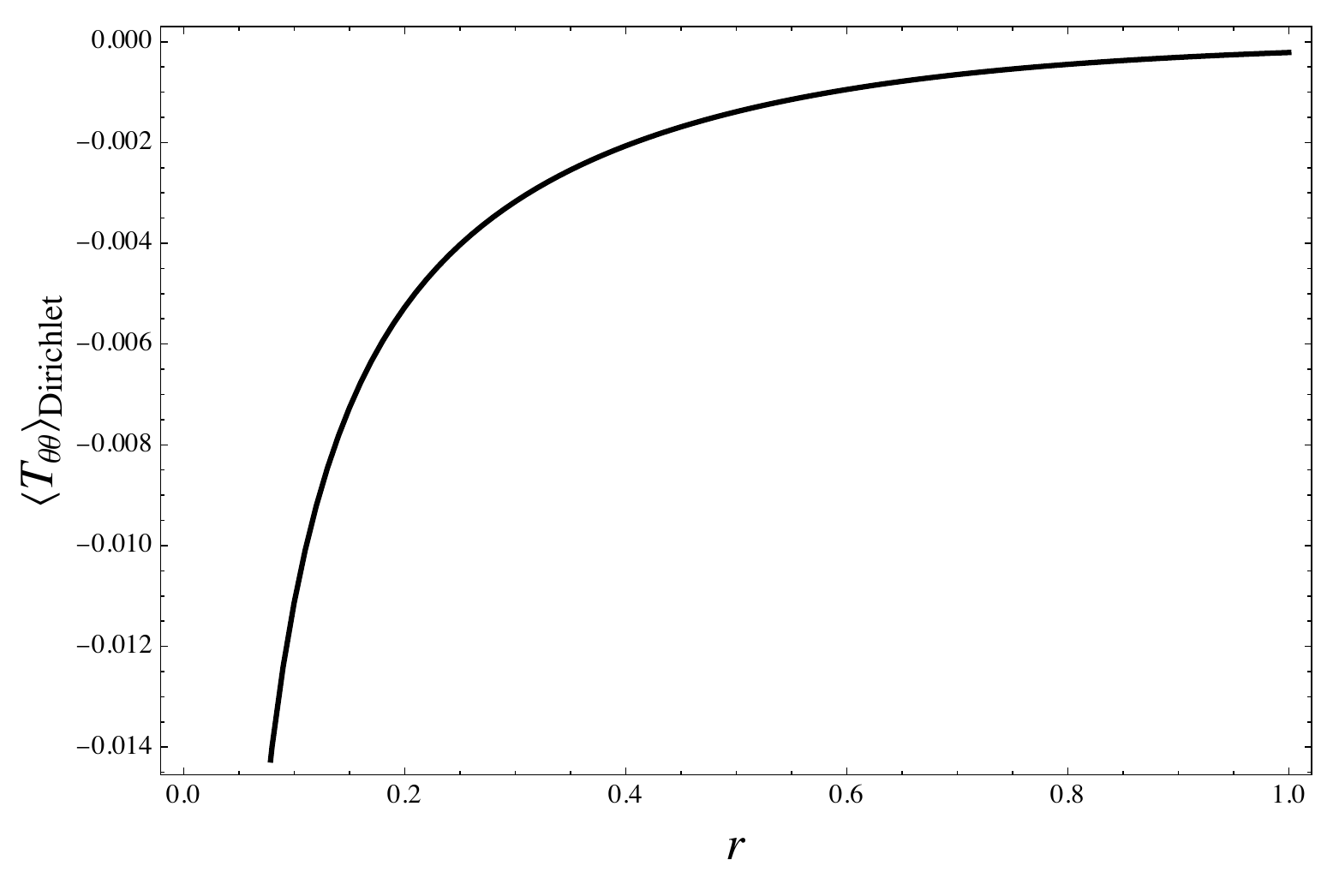}
\caption{Renormalized stress-energy tensor components under Dirichlet boundary conditions for $\alpha=0.9$ and $M=2$. From top to bottom: $\langle T_{tt}\rangle_{\text{Dirichlet}}$, $\langle T_{rr}\rangle_{\text{Dirichlet}}$, and $\langle T_{\theta\theta}\rangle_{\text{Dirichlet}}$ as functions of the radial coordinate $r$.}
\label{fig2}
\end{figure}

Notice that $G^+_\textrm{Dirichlet}(\mathsf{x},\mathsf{x}')$ in Eq.~(\ref{Grenorm}) (for $\alpha>1/2$) can be split in two terms. The first term is given by
\begin{equation}
    W^{(1)}_{\textrm{Dirichlet}}=\frac{1}{4\pi\sqrt{2\sigma_0}}\left(M\sqrt{2\sigma_0}+\frac{M^3}{3!}(2\sigma_0)^{3/2}+\cdots\right)
\end{equation}
so that $\langle T^{(1)}_{\mu\nu}\rangle_{\textrm{Dirichlet}}$ can be treated analytically. 
We have
\begin{equation}\begin{aligned}
\langle T^{(1)}_{00}\rangle_{\textrm{Dirichlet}}&=-\frac{M^3}{12\pi}+\Theta_{00},\\
\langle T^{(1)}_{11}\rangle_{\textrm{Dirichlet}}&=\frac{M^3}{12\pi}+\Theta_{11},\\
\langle T^{(1)}_{22}\rangle_{\textrm{Dirichlet}} &=r^2\alpha^2\frac{M^3}{12\pi}+\Theta_{22}.
\end{aligned}
\end{equation}
The off-diagonal components will be null, given that $t\to -t$ and $\theta\to-\theta$ are symmetries of the spacetime; hence $\langle T_{tr}\rangle_{\textrm{Dirichlet}}=\langle T_{-tr}\rangle_{\textrm{Dirichlet}}=\frac{\partial (-t)}{\partial x^\mu}\frac{\partial r}{\partial x^\mu}\langle T_{\mu\nu}\rangle_{\textrm{Dirichlet}}=-\langle T_{t r}\rangle_{\textrm{Dirichlet}}\Rightarrow\langle T_{tr}\rangle_{\textrm{Dirichlet}}=0$ and the same argument applies to the other non-diagonal components.
Thus we see that $\langle T_{\mu\nu}\rangle=\frac{M^3}{12\pi}g_{\mu\nu}+\Theta_{\mu\nu}$ and, since we expect to recover Minkowski when $\alpha\to 1$, we choose $\Theta_{\mu\nu}=-\frac{M^3}{12\pi}g_{\mu\nu}$. We then arrive at
\begin{equation}
\langle T_{\mu\nu}\rangle_{\textrm{Dirichlet}}=-W_{\textrm{Dirichlet\,}\mu\nu}^{(2)}+\frac{1}{2}W^{(2)}_{\textrm{Dirichlet\,};\mu\nu}-\frac{1}{4}g_{\mu\nu}\square W^{(2)}_{\textrm{Dirichlet}},
\label{tmunu2}
\end{equation}
with
\begin{equation}
W^{(2)}_{\textrm{Dirichlet}}(\mathsf{x},\mathsf{x}')=-\frac{1}{8\pi^2\alpha}\sum_{j=\pm}\int_0^\infty\textrm{d}\xi\,\frac{\sin\!\left[\tfrac{j(\alpha \Delta\theta+\pi)}{\alpha}\right]\tfrac{e^{-M\sqrt{2\sigma_\xi}}}{\sqrt{2\sigma_\xi}}}{\cosh\!\left(\tfrac{\xi}{\alpha}\right)-\cos\!\left(\tfrac{j\alpha\Delta\theta+\pi}{\alpha}\right)}.
\label{w2}
\end{equation}
We first differentiate Eq.~(\ref{w2}) under the integral symbol according to Eq.~(\ref{tmunu2}) and integrate the resulting expression numerically. For $\alpha=0.9$ and $M=2$ we obtain $\langle T_{\mu\nu}\rangle_{\textrm{Dirichlet}}$ as a function of $r$ as shown in Fig.~\ref{fig2}. We verified numerically that 
$\nabla^\mu\langle T_{\mu\nu}\rangle_{\textrm{Dirichlet}}=0$. 
This result is consistent with the expectation that covariant 
conservation should hold when Dirichlet boundary conditions are imposed.

There remain two additional contributions to 
$\langle T_{\mu\nu}\rangle$, namely 
$\langle T_{\mu\nu}\rangle_{\textrm{bc}}$ and 
$\langle T_{\mu\nu}\rangle_{\textrm{bound}}$. 
For the bound-state sector, we obtain
\begin{equation}
\langle T_{\mu\nu}\rangle_{\textrm{bound}}
=-W^{\textrm{bound}}_{\mu\nu}
+\tfrac{1}{2}W^{\textrm{bound}}_{;\mu\nu}
-\tfrac{1}{4}g_{\mu\nu}\,\square W^{\textrm{bound}},
\end{equation}
with 
\begin{equation}
W^{\textrm{bound}}(\mathsf{x},\mathsf{x}')
= \frac{q^2}{2\pi \alpha}\,
\frac{K_0(q r)\,K_0(q r')}
{\sqrt{M^2-q^2}}\,
e^{-i\sqrt{M^2-q^2}(t-t')}.
\end{equation}
Explicit evaluation then yields
\begin{equation}
\begin{aligned}
\langle T_{tt}\rangle_{\textrm{bound}} 
&= \frac{\big(2M^2q^2 - q^4\big) K_0(qr)^2 
+ q^4 K_1(qr)^2}
{4\pi \alpha \sqrt{M^2-q^2}}, \\[6pt]
\langle T_{rr}\rangle_{\textrm{bound}} 
&= \frac{q^4 \big(K_1(qr)^2 - K_0(qr)^2\big)}
{4\pi \alpha \sqrt{M^2-q^2}}, \\[6pt]
\langle T_{\theta\theta}\rangle_{\textrm{bound}} 
&= -\frac{\alpha q^4 r^2 \big(K_0(qr)^2 + K_1(qr)^2\big)}
{4\pi \sqrt{M^2-q^2}}.
\end{aligned}
\label{tbound}
\end{equation}
Finally, from Eq.~(\ref{tbound}) it follows by direct computation that 
\begin{equation}
\nabla^\mu \langle T_{\mu\nu}\rangle_{\textrm{bound}} = 0,
\end{equation}
confirming the covariant conservation of the bound-state contribution. 

Now we consider
\begin{equation}
\begin{aligned}
W_{\textrm{bc}}(\mathsf{x},\mathsf{x}')
&=\frac{1}{4\pi\alpha}\int_0^\infty \mathrm{d}\lambda \,
\frac{\lambda}{\omega}\,\frac{\beta(\lambda)}{1+\beta^2(\lambda)} \\
&\quad \times \Bigg\{\beta(\lambda)\Big[-J_0(\lambda r)J_0(\lambda r')
+Y_0(\lambda r)Y_0(\lambda r')\Big] 
\\&\quad+ \Big[J_0(\lambda r)J_0(\lambda r')
+Y_0(\lambda r)Y_0(\lambda r')\Big]\Bigg\}
e^{-i\Delta t}.
\end{aligned}
\end{equation}
The corresponding stress-energy contribution is
\begin{equation}
\langle T_{\mu\nu}\rangle_{\textrm{bc}}
= -W_{\textrm{bc}\,\mu\nu}
+ \tfrac{1}{2}W_{\textrm{bc};\mu\nu}
- \tfrac{1}{4}g_{\mu\nu}\,\square W_{\textrm{bc}}.
\end{equation}

\begin{figure}[htb!]
    \centering
    \includegraphics[scale=0.45]{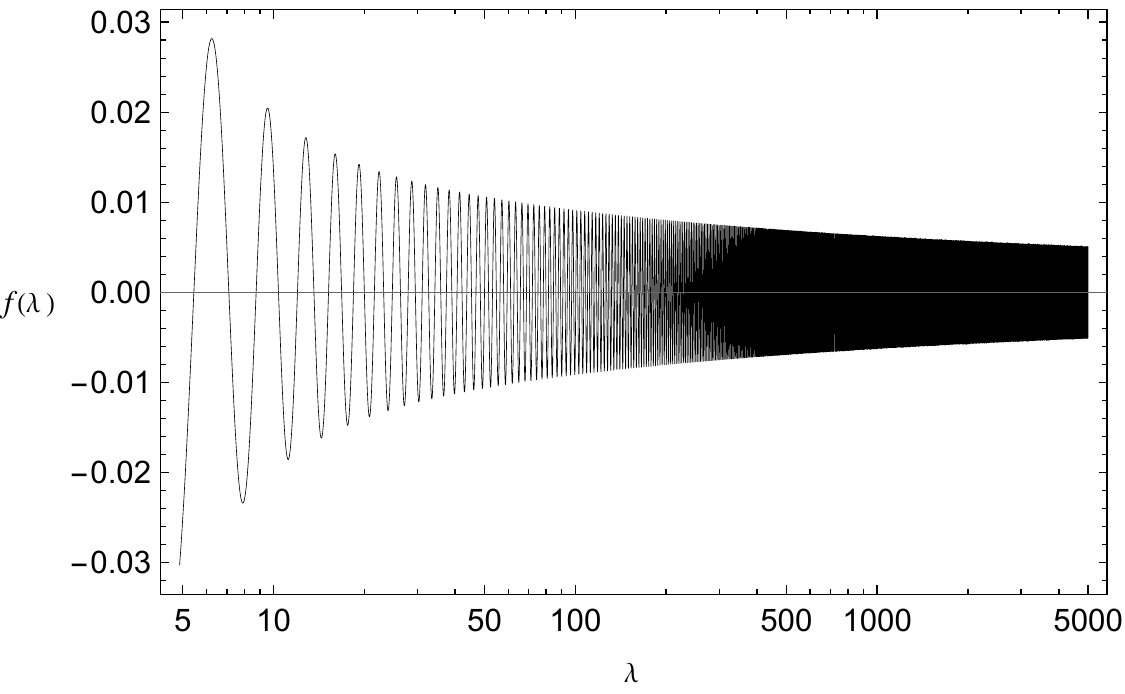}
    \caption{Integrand used in the calculation of 
$\langle T_{00}\rangle_{\textrm{bc}}$. 
The explicit expression for $f(\lambda)$ is provided in Appendix \ref{appendix:numerics}. 
Note that $f(\lambda)$ tends to zero as $\lambda \to \infty$, albeit with very slow convergence.}
    \label{fig3}
\end{figure}

Since no closed analytic form is available for 
$W^{\textrm{bc}}(\mathsf{x},\mathsf{x}')$, 
we adopt the following strategy: first we differentiate under the integral 
sign, and then integrate the resulting expression. 
This procedure, however, comes with a caveat. 
As shown in Fig.~\ref{fig3}, the convergence of the integral is 
extremely slow when evaluating 
$\langle T_{tt}\rangle_{\textrm{bc}}$ and 
$\langle T_{rr}\rangle_{\textrm{bc}}$, 
while in the case of 
$\langle T_{\theta\theta}\rangle_{\textrm{bc}}$ 
the integral actually diverges.

To handle the slow convergence of the integrals appearing in the calculation of $\langle T_{00}\rangle_{\textrm{bc}}$, we first expand the integrand at large $\lambda$ and extract its asymptotic form. This asymptotic contribution is then treated analytically using integration by parts, which isolates rapidly decaying terms and leaves a remainder that converges much faster. The resulting expression allows for stable numerical evaluation with only a weak dependence on the cutoff parameter once it is sufficiently large. The same strategy applies to $\langle T_{rr}\rangle_{\textrm{bc}}$, while in the case of $\langle T_{\theta\theta}\rangle_{\textrm{bc}}$ this direct approach fails due to the divergence of the resulting integral (changing the order of operations of integration and differentiation is illegal in this case).  In this situation, we exploit the conservation of the stress-energy tensor to express $\langle T_{\theta\theta}\rangle_{\textrm{bc}}$ in terms of $\langle T_{rr}\rangle_{\textrm{bc}}$, which can be computed numerically and differentiated after interpolation. The full details of this procedure are presented in Appendix \ref{appendix:numerics}. In Fig.~\ref{tfinal} we plot $\langle T_{\mu\nu}\rangle$, together with 
$\langle T_{\mu\nu}\rangle_{\textrm{Dirichlet}}$, to compare their 
orders of magnitude.
\begin{figure}[htb!]
\centering

    \includegraphics[width=\linewidth]{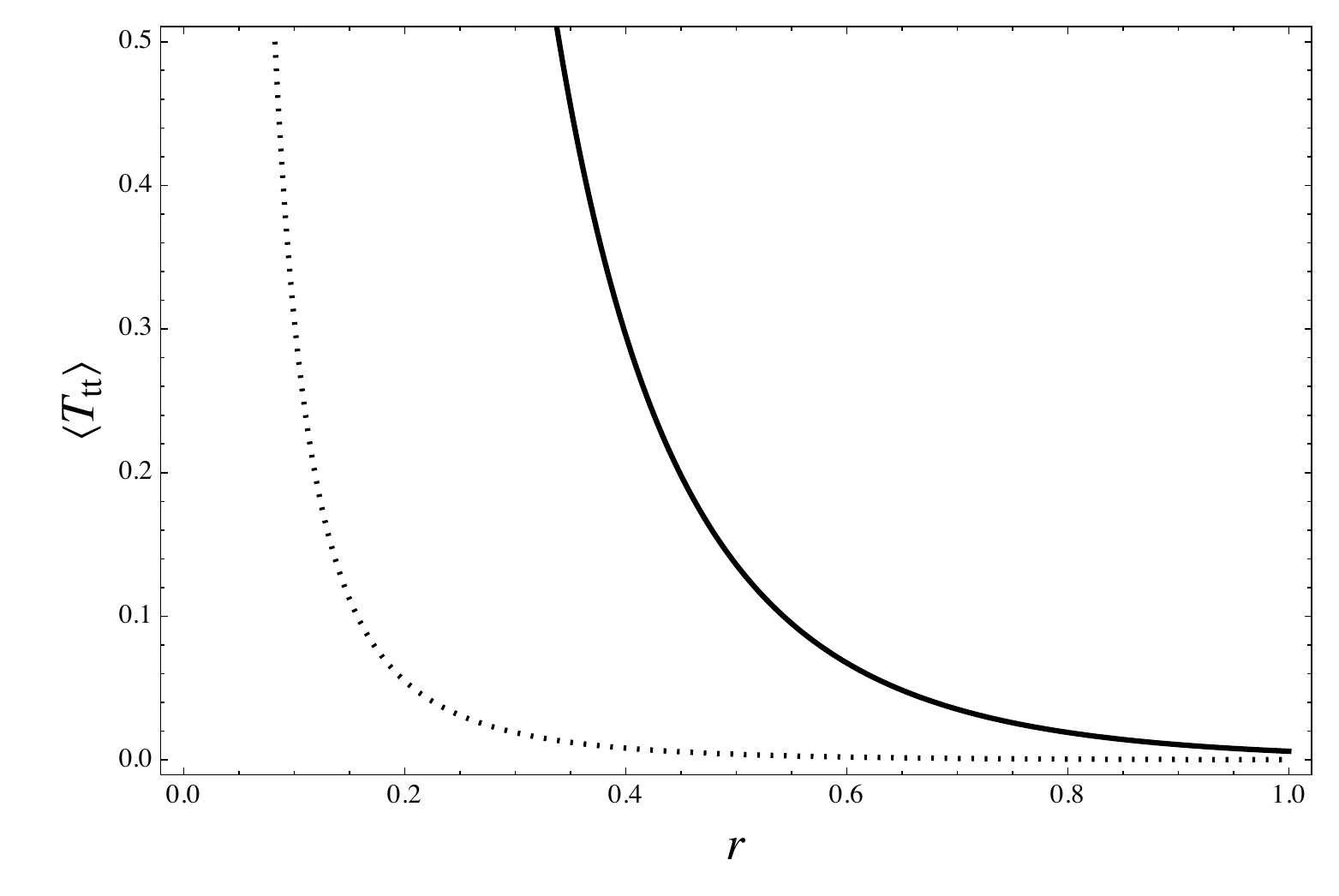}

    \includegraphics[width=\linewidth]{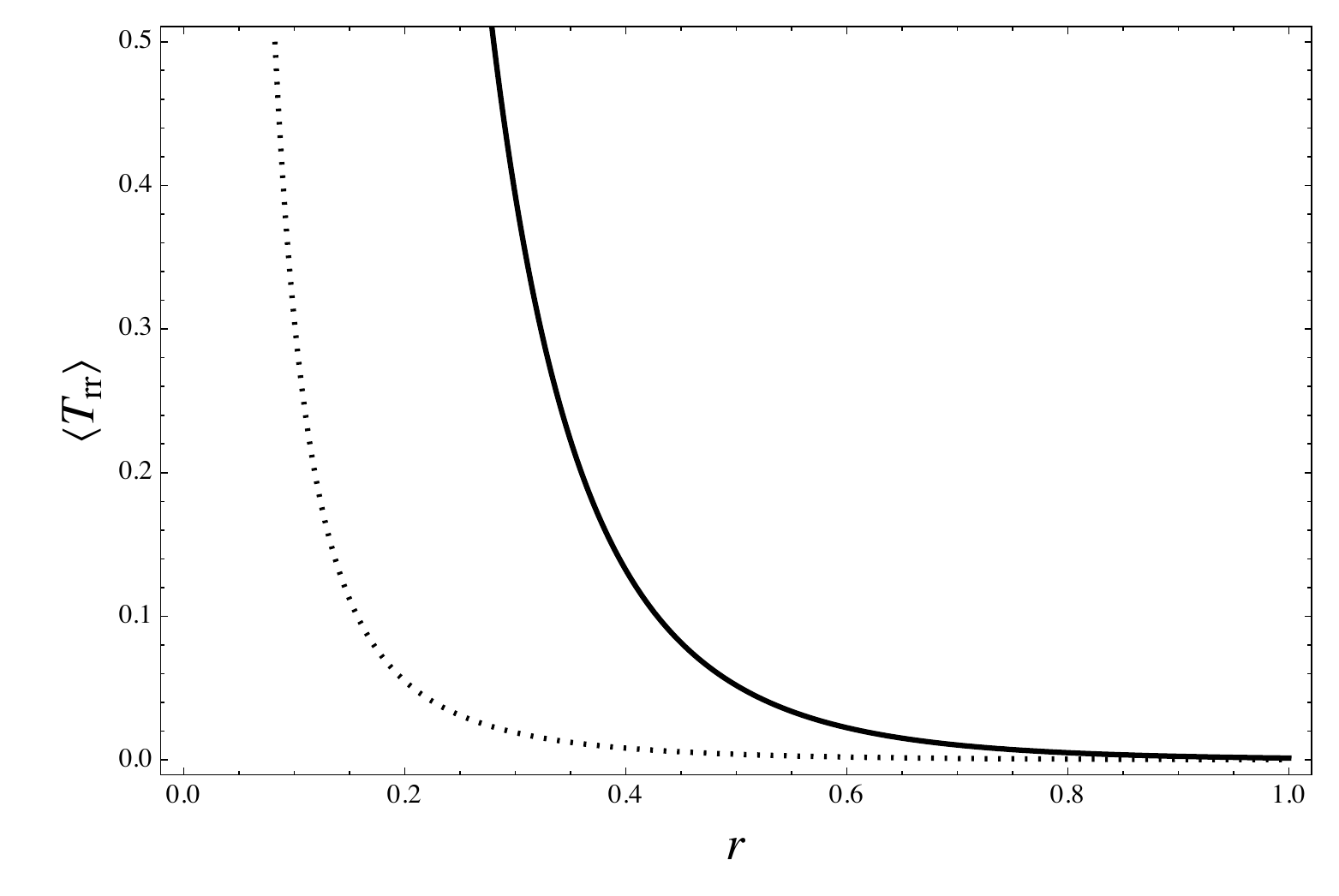}
   
    \includegraphics[width=\linewidth]{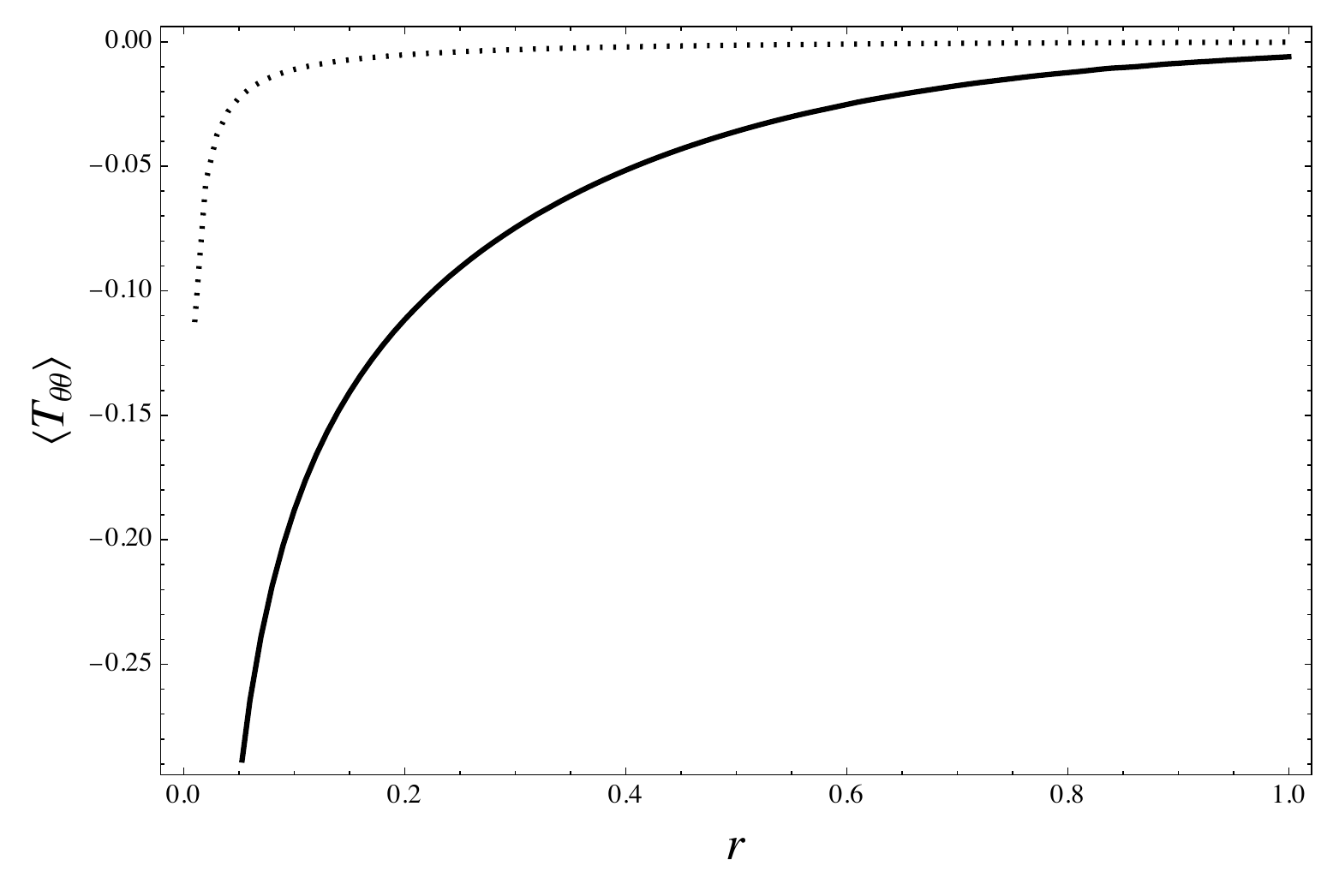}
   
\caption{From top to bottom: 
$\langle T_{tt}\rangle_{\text{Dirichlet}}$ (dotted line) and $\langle T_{tt}\rangle$ (solid line), 
$\langle T_{rr}\rangle_{\text{Dirichlet}}$ (dotted line) and $\langle T_{rr}\rangle$ (solid line), 
$\langle T_{\theta\theta}\rangle_{\text{Dirichlet}}$ (dotted line) and $\langle T_{\theta\theta}\rangle$ (solid line), 
as functions of the radial coordinate $r$. 
The boundary-condition contribution clearly dominates over the Dirichlet one.}
\label{tfinal}
\end{figure}

\section{Concluding Remarks}
\label{concluding remarks}
We have computed the vacuum fluctuations and the renormalized stress-energy tensor of a massive scalar field in a conical spacetime, subject to a nontrivial boundary condition parametrized by $q$ at the apex, under the stability condition $M>q$. The existence of a discrete square-integrable mode enables a natural interpretation of this field as a covariant model of an extended particle detector. The coupling of such a detector to a test field may be described by an interaction Lagrangian of the form
\begin{equation}
    \mathcal{L}_{\textrm{int}}=\lambda\,\zeta(\mathsf{x})\,\Psi(\mathsf{x})\,\phi_\textrm{test}(\mathsf{x}),
\end{equation}
where $\zeta(\mathsf{x})$ is a profile function and $|\lambda|\ll 1$.  

Furthermore, by inserting the renormalized stress-energy tensor associated with $\Psi$ into Einstein’s equations, one can investigate how the detector backreacts on the surrounding geometry. In addition, this framework allows for couplings to linearized gravity through
\begin{equation}
    \bar{\mathcal{L}}_\textrm{int}=\langle T_{\mu\nu} \rangle h^{\mu\nu}.
\end{equation}
We plan to explore both of these directions in future work.

\section*{Acknowledgments}

J. P. M. P. is thankful for the support provided in part by Conselho Nacional de Desenvolvimento Científico e Tecnológico (CNPq, Brazil), Grant No. 305194/2025-9. R.A.M. was partially supported by Conselho Nacional de Desenvolvimento Cient\'{\i}fico e Tecnol\'{o}gico (CNPq, Brazil) under Grant No. 316780/2023-5. V.H.M.R. acknowledges the financial support of Coordenação de Aperfeiçoamento de Pessoal de Nível Superior (CAPES) - Brazil, Finance Code 001, and gratefully acknowledges the kind hospitality of the Department of Mathematics of the University of Genova.

\appendix
\section{DETAILS OF THE NUMERICAL PROCEDURE}
\label{appendix:numerics}

In this appendix we present the detailed procedure used to handle the 
slowly convergent integrals that arise in the evaluation of 
$\langle T_{tt}\rangle_{\textrm{bc}}$ and $\langle T_{rr}\rangle_{\textrm{bc}}$. 
For concreteness, we illustrate our numerical method in the calculation of 
$\langle T_{tt}\rangle_{\textrm{bc}}$, while emphasizing that the same 
strategy applies to $\langle T_{rr}\rangle_{\textrm{bc}}$.

We first consider the integrand of $\langle T_{tt}\rangle_{\textrm{bc}}$, i.e., 
\begin{equation}
    \langle T_{tt}\rangle_{\textrm{bc}}=\int_0^\infty{f(\lambda)d\lambda},
\end{equation}
with
\begin{widetext}
\begin{equation}
\begin{aligned}
f(\lambda) &= \frac{1}{8\alpha \sqrt{M^2+\lambda^2}\,\big(4\log^2\!\tfrac{q}{\lambda}+\pi^2\big)}
\Bigg\{\lambda \Big[
4(2M^2+\lambda^2)\log\!\left(\tfrac{q}{\lambda}\right)J_0(\lambda r)Y_0(\lambda r) 
- \pi (2M^2+\lambda^2) J_0(\lambda r)^2 \\
&\quad + \pi (2M^2+\lambda^2) Y_0(\lambda r)^2
+ 4\lambda^2 \log\!\left(\tfrac{q}{\lambda}\right) J_1(\lambda r)Y_1(\lambda r) 
- \pi \lambda^2 J_1(\lambda r)^2 
+ \pi \lambda^2 Y_1(\lambda r)^2
\Big]\Bigg\}.
\end{aligned}
\end{equation}
\end{widetext}
Then we expand $f(\lambda)$ around $\lambda= +\infty$ and obtain
\begin{widetext}
\begin{equation}
\begin{aligned}
f_{\textrm{asymp}}(p) 
&=\frac{1}{256 \pi \alpha p^2 r^3 \sqrt{M^2+p^2}\,\big(4\log^2\!\tfrac{q}{p}+\pi^2\big)} 
\Bigg\{4\cos(2pr)\Big[\big(-64M^2p^2r^2+M^2-4p^2\big)\log\!\tfrac{q}{p} 
+ 8\pi pr (M^2+2p^2)\Big] \\
&\quad -2\sin(2pr)\Big[32pr(M^2+2p^2)\log\!\tfrac{q}{p} 
+ \pi M^2(64p^2r^2-1)+4\pi p^2\Big]\Bigg\} \\
&\equiv u(p)+v(p),
\end{aligned}
\end{equation}
\end{widetext}
where
\begin{widetext}
\begin{equation}
\begin{aligned}
u(p) &= \left[
\frac{\big((M^2(1-64p^2r^2)-4p^2)\log\!\tfrac{q}{p}
+8\pi pr(M^2+2p^2)\big)}
{64\pi \alpha p^2 r^3 \sqrt{M^2+p^2}\,\sqrt{4\log^2\!\tfrac{q}{p}+\pi^2}}
\cos(2pr)\right]  \times      \frac{1}{\sqrt{4\log^2\!\tfrac{q}{p}+\pi^2}}
\equiv g_1(p)h(p),\\
v(p) &= -\left[
\frac{\big(32pr(M^2+2p^2)\log\!\tfrac{q}{p}
+\pi M^2(64p^2r^2-1)+4\pi p^2\big)}
{128\pi \alpha p^2 r^3 \sqrt{M^2+p^2}\,\sqrt{4\log^2\!\tfrac{q}{p}+\pi^2}}
\sin(2pr)\right]\times \frac{1}{\sqrt{4\log^2\!\tfrac{q}{p}+\pi^2}}
\equiv g_2(p)h(p),
\end{aligned}
\end{equation}
\end{widetext}
with
\[
h(p)=\frac{1}{\sqrt{\pi^2+\log^2\!\tfrac{q}{p}}}.
\]
It can be verified that the bracketed terms, namely $g_1(p)$ and $g_2(p)$, 
satisfy
\[
\Big|\int_\Lambda^x g_i(\lambda)\,\mathrm{d}\lambda\Big|\leq M_i,
\quad \forall x\geq \Lambda, \quad i=1,2,
\]
while $h'(p)\leq 0$ and $\lim_{p\to\infty}h(p)=0$. 
Hence, by Dirichlet’s test for improper integrals, 
$f_{\textrm{asymp}}(p)$ converges, as does the original integral of $f(p)$.

Proceeding in this way we rewrite
\begin{equation}
\begin{aligned}
   \int_0^\infty f(\lambda)\,\mathrm{d}\lambda 
   &= \int_0^\Lambda f(\lambda)\,\mathrm{d}\lambda
   + \int_\Lambda^\infty f_{\textrm{asymp}}(\lambda)\,\mathrm{d}\lambda \\
   &\quad + \int_\Lambda^{\infty}\!\Big[f(\lambda)-f_{\textrm{asymp}}(\lambda)\Big]\,\mathrm{d}\lambda.
   \label{integrals}
\end{aligned}
\end{equation}
Now we define $l_1(p)$ and $l_2(p)$ by
\begin{equation}\begin{aligned}
    u(p)&=l_1(p) \cos{(2pr)},\\
    v(p)&=l_2(p) \sin{(2p r).}
\end{aligned}\end{equation}
The second term on the right-hand side of Eq.~(\ref{integrals}) is then integrated by parts, yielding
\begin{equation}
    \begin{aligned}
\int_\Lambda^\infty f_{\textrm{asymp}}(\lambda)\,\mathrm{d}\lambda
&= l_2(\Lambda)\,\frac{\cos(2\Lambda r)}{2r}
   - l_1(\Lambda)\,\frac{\sin(2\Lambda r)}{2r}
   + R(\Lambda).
\end{aligned}
\end{equation}
This procedure isolates rapidly decaying contributions and leaves a remainder 
$R(\Lambda)+\int_\Lambda^{\infty}[f(\lambda)-f_{\textrm{asymp}}(\lambda)]\,\mathrm{d}\lambda$ 
that converges extremely fast to zero, thereby ensuring numerical stability and 
only a weak dependence on the cutoff $\Lambda$ at large values. 

The same strategy applies to $\langle T_{rr}\rangle_{\textrm{bc}}$, 
whereas for $\langle T_{\theta\theta}\rangle_{\textrm{bc}}$ the divergence of 
the integrand requires exploiting the conservation of the stress-energy tensor. 
In this case, $\langle T_{\theta\theta}\rangle_{\textrm{bc}}$ was obtained from 
$\langle T_{rr}\rangle_{\textrm{bc}}$ through the $\nu=1$ component of 
$\nabla^\mu T_{\mu\nu}=0$, namely
\begin{equation}
    \langle T_{\theta\theta}\rangle_{\textrm{bc}}
    = \alpha^2 r^3 \left(
    \frac{\partial \langle T_{rr}\rangle_{\textrm{bc}}}{\partial r}
    + \frac{1}{r}\,\langle T_{rr}\rangle_{\textrm{bc}}
    \right).
\end{equation}

The radial derivative in $\langle T_{\theta\theta} \rangle_{\textrm{bc}}$ was computed numerically using an adaptive Savitzky-Golay differentiator.

\end{document}